\documentclass[aps,twocolumn,superscriptaddress,altaffilletter,lengthcheck,
tightenlines,showpacs,showkeys]{revtex4}
\usepackage{amsmath} 
\usepackage{amssymb} 

\newcommand{\ben}{\begin{eqnarray}}
\newcommand{\een}{\end{eqnarray}}
\newcommand{\be}{\begin{equation}}
\newcommand{\ee}{\end{equation}}
\newcommand{\ba}{\begin{eqnarray}}
\newcommand{\ea}{\end{eqnarray}}
\newcommand{\n}{\label}

\newcommand{\ga}{\gamma}
\newcommand{\ro}{\rho}

\newcommand{\Om}{\Omega}

\newcommand{\bn}{\begin{equation}\label}

\usepackage[dvipdf]{epsfig}
\usepackage{color}

\begin{document}

\title{Interacting dark sector with transversal interaction}

%%%%%%%%%%%%%%%%%%%%%%%%%%%%%%%%%%%%%%%%%%%%%%%%%%%%%%%%%%%%%%%%%%%%%%%%%%%%%%%%%%%%%%%%%%%%%%%%%%%%
\author{Luis P. Chimento}%\email{chimento@df.uba.ar}
\affiliation{Departamento de F\'{\i}sica, Facultad de Ciencias Exactas y Naturales,  Universidad de Buenos Aires and IFIBA, CONICET, Ciudad Universitaria, Pabell\'on I, Buenos Aires 1428 , Argentina}
\author{Mart\'{\i}n G. Richarte}%\email{martin@df.uba.ar}
\affiliation{Departamento de F\'{\i}sica, Facultad de Ciencias Exactas y Naturales,  Universidad de Buenos Aires and IFIBA, CONICET, Ciudad Universitaria, Pabell\'on I, Buenos Aires 1428 , Argentina}
%%%%%%%%%%%%%%%%%%%%%%%%%%%%%%%%%%%%%%%%%%%%%%%%%%%%%%%%%%%%%%%%%%%%%%%%%%%%%%%%%%%%%%%%%%%%%%%%%%%%

%Departamento de Física, Facultad de Ciencias Exactas y Naturales, Universidad de Buenos Aires and IFIBA, CONICET, Cuidad Universitaria, Buenos Aires 1428, Argentina

%\date{\today}
\bibliographystyle{plain}

\begin{abstract}

We investigate the interacting dark sector composed of  dark matter, dark energy,  and dark radiation for a  spatially flat Friedmann-Robertson-Walker (FRW) background by introducing
a three-dimensional internal space spanned by the interaction vector  $\mathbf{Q}$  and solve the source equation for a linear transversal interaction. Then, we explore a realistic model with dark matter coupled to a scalar field  plus a decoupled radiation term, analyze the amount of dark energy in the radiation era and find that our model is consistent with the recent  measurements  of  cosmic microwave background anisotropy coming from Planck  along with   the future constraints achievable by CMBPol experiment.
\end{abstract} 
\vskip 1cm

%\pacs{}

%\date{\today}
\bibliographystyle{plain}

\maketitle

%%%%%%%%%%%%%%%%%%%%%%%%%%%%%%%%%
\section{Introduction}
%%%%%%%%%%%%%%%%%%%%%%%%%%%%%%%
One could consider an exchange of energy between the components of the dark sector, so  dark matter not only can feel the presence of   dark energy through a gravitational expansion of the universe but also can interact between them \cite{jefe1}, \cite{T1}, \cite{jefe4}, \cite{ot}.  A coupling between dark energy and dark matter changes the background evolution of the dark sector allowing us to constrain any particular kind of interaction and giving  rise to a richer cosmological dynamics compared with non-interacting models \cite{jefe1}, \cite{NQ}. One way to extend the insight about the dark matter-dark energy interacting mechanism is to explore a bigger picture in which a third component is added, perhaps a  weakly interacting radiation term as it occurs within warm inflation paradigm. A step forward for constraining dark matter and dark energy  is to use  the physic behind  recombination or big-bang nucleosynthesis epochs by adding a decoupled radiation term to the dark sector for taking into account the stringent bounds  related to  the behavior of dark energy at early times 
 \cite{jefe4}, \cite{hmi1}, \cite{hmi2}. The authors have explored the behavior of a universe filled with  three interacting components, i.e., dark energy, dark matter, and dark radiation \cite{jefe2} plus a decoupled radiation term (Model I). Then, they have extended the aforesaid scheme by considering a more realistic model with dark matter coupled to a scalar field plus a decoupled radiation (Model II) \cite{jefe3}.  We perform a cosmological constraint using the updated Hubble data \cite{H29},  bounds for  early dark energy \cite{EDE1}, \cite{Amendola},  \cite{Cyburt}, \cite{Cyburt2} and compare  our  constraints on cosmological parameters with the bounds reported by Planck  measurements \cite{Planck2013} and WMAP-9 project \cite{WMAP9} .
%%%%%%%%%%%%%%%%%%%%%%%%%%%%%%%%%%%%%%%%%%%
\section{Model I: Enlarged dark sector }
%%%%%%%%%%%%%%%%%%%%%%%%%%%%%%%%%%%%%%%%%%%
We consider a spatially flat FRW spacetime filled with three interacting fluids, namely, dark energy, dark matter and radiation so that  the dynamics of the model is given by  the Friedmann and conservation equations, 
\be
\n{01}
3H^{2}=\ro=\ro_{x}+ \ro_{m}+\ro_{r},
\ee
\be
\n{02}
\dot{\ro}+3H(\ro_{x}+p_{x}+\ro_{m}+p_{m}+\ro_{r}+p_{r})=0,
\ee
where $H = \dot a/a$ is the Hubble expansion rate and  $a$ is the scale factor. Introducing the variable $\eta = \ln(a/a_0)^{3}$, with $a_0$ the present value of the scale factor and $' \equiv d/d\eta$ and assuming linear barotropic equations of state for the fluids, $p_{i}=(\gamma_{i}-1)\ro_{i}$, with constant barotropic indexes $\ga_i$ being $i={x,m,r}$, Eq. (\ref{02}) can be recast 
\be
\n{03}
\ro'=-\gamma_{x}\ro_{x}-\gamma_{m}\ro_{m}- \gamma_{r}\ro_{r},
\ee
where the $\ga_i$ satisfy the conditions $0<\ga_x<\ga_m<\ga_r$.  The interaction terms $3H Q_{x}(\ro,\ro',\eta)$, $3H Q_{m}(\ro,\ro',\eta)$ and $3H Q_{r}(\ro,\ro',\eta)$ between the components are introduced by  splitting (\ref{03}) into three equations:
\ben
\n{04}
\ro_x' + \ga_{x} \ro_x =  Q_{x},
\\
\n{05}
\ro_m' + \ga_{m} \ro_m = Q_{m},
\\
\n{06}
\ro_r' + \ga_{r} \ro_r = Q_{r},
\een
where $ Q_{x}+  Q_{m}+ Q_{r}=0$ to recover the whole conservation equation (\ref{03}).

\begin{figure}[h!]
\begin{center}
\includegraphics[height=6.9cm,width=7.5cm]{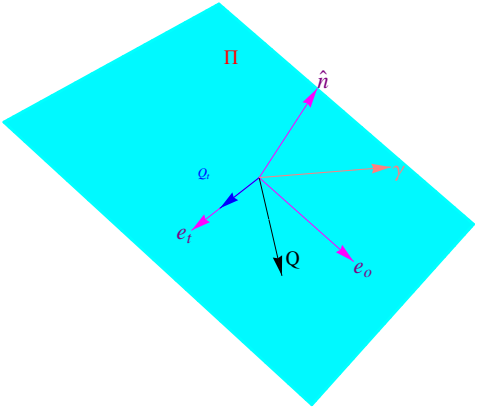}
\caption{The plot shows the vector basis $\{\mathbf{e_t},\mathbf{e_o},\mathbf{n}\}$, the interaction plane $Q_{x}+Q_{m}+Q_{r}=0$, the interaction vector  $\mathbf{Q}=Q_t\,\mathbf{e_t}+Q_o\,\mathbf{e_o}$ and the vector $\mbox{\boldmath ${\gamma}$}=\ga_n\mathbf{n}+\ga_o\mathbf{e_o}$.}
\label{F1}
\end{center}
\end{figure}

We introduce a 3-dimensional internal space where the three interaction terms and barotropic indexes are viewed as vectors. The interaction vector $\mathbf{Q}=(Q_{x}, Q_{m}, Q_{r})$ belongs to the interaction plane $ \Pi$ generated by $ Q_{x}+  Q_{m}+ Q_{r}=0$ while the barotropic vector $\mbox{\boldmath ${\gamma}$}=(\ga_x,\ga_m,\ga_r)$ intersects it. At this intersection, we set the  coordinate  system origin of the internal space see Fig. (\ref{F1}). 

After differentiating Eq. (\ref{03}) and using Eqs. (\ref{04})-(\ref{06}) we have  the additional equation 
\be
\n{07}
\ro''+\mbox{\boldmath ${\gamma}$}\cdot\mathbf{Q}=\ga^{2}_{x}\ro_{x}+\ga^{2}_{m}\ro_{m}+ \ga^{2}_{r}\ro_{r}.
\ee
Solving the algebraic system of equations in the $(\ro_{x}, \ro_{m}, \ro_{r})$ variables (\ref{01}), (\ref{03}), and (\ref{07}), we obtain 
\be
\n{08}
\ro_x=\frac{\ga_{m}-\ga_{r}}{\Delta}\left[ \ga_{m}\ga_{r}\ro+(\ga_{m}+\ga_{r})\ro'+ \ro''+\mbox{\boldmath ${\gamma}$}\cdot\mathbf{Q}\right],
\ee
\be
\n{09}
\ro_m=\frac{\ga_{r}-\ga_{x}}{\Delta}\left[ \ga_{x}\ga_{r}\ro+ (\ga_{x}+\ga_{r})\ro'+ \ro''+\mbox{\boldmath ${\gamma}$}\cdot\mathbf{Q}\right],
\ee
\be
\n{10}
\ro_r=\frac{\ga_{x}-\ga_{m}}{\Delta}\left[ \ga_{x}\ga_{m}\ro+ (\ga_{x}+\ga_{m})\ro'+\ro''+\mbox{\boldmath ${\gamma}$}\cdot\mathbf{Q}\right],
\ee
where $\Delta=(\ga_{x}-\ga_{m}).(\ga_{x}-\ga_{r}).(\ga_{m}-\ga_{r})$ is the determinant of that system of equations.  Equations (\ref{08})-(\ref{10}) represent the straightforward extension of the interacting two-fluid scenario studied in \cite{jefe1}. Following Ref. \cite{jefe1},  we replace Eqs. (\ref{08})-(\ref{10}) into (\ref{04})-(\ref{06}) and obtain the third order differential equation, that we call ``source equation'', for the total energy density; 
$$
\ro'''+(\ga_{x}+\ga_{m}+\ga_{r})\ro''+(\ga_{x}\ga_{r}+\ga_{x}\ga_{m}+\ga_{m}\ga_{r})\ro'+
$$
\be
\n{11}
\ga_{x}\ga_{m}\ga_{r}\ro=\ga_{x}Q_{m}\ga_{r}+ \ga_{r}Q_{x}\ga_{m}+\ga_{m}Q_{r}\ga_{x}-\mbox{\boldmath ${\gamma}$}\cdot\mathbf{Q'}.
\ee
To simplify the mathematical treatment of the model, we exploit the internal space structure. To this end, we introduce a pair of vectors $\{\mathbf{e_t},\mathbf{e_o}\}$ in $\Pi$, $\mathbf{e_t}$ is orthogonal to the vector $\mbox{\boldmath ${\gamma}$}$, so $\mbox{\boldmath ${\gamma}$}\cdot\mathbf{e_t}=0$, the orthogonal projection of the vector $\mbox{\boldmath ${\gamma}$}$ on  $\Pi$ defines the direction of the vector $\mathbf{e_o}$, and a normal vector $\mathbf{n}=(1,1,1)$ to $\Pi$.

Therefore, $\mathbf{Q}=Q_t\,\mathbf{e_t}+Q_o\,\mathbf{e_o}$, where $Q_t$ and $Q_o$ are the components of the interaction vector $\mathbf{Q}$ on $\Pi$ and $\mathbf{n}\cdot \mathbf{Q}=0$.
Due that the direction of $\mathbf{e_t}$ is the unique with the property of being orthogonal to $\mbox{\boldmath ${\gamma}$}$, we adopt this property as a simple criteria for selecting only those interactions which are collinear with the aforesaid preferred direction in $\Pi$ that we call ``transversal interaction'', so 
\bn{qt}
\mathbf{Q_t}=Q_t\,(\ga_{m}-\ga_{r}, \ga_{r}-\ga_{x}, \ga_{x}-\ga_{m}),
\ee
ensuring that $\mbox{\boldmath ${\gamma}$}\cdot \mathbf{Q}=0$. The transversal character of the interaction vector (\ref{qt}) removes the terms $\mbox{\boldmath ${\gamma}$}\cdot\mathbf{Q}$ and $\mbox{\boldmath${\gamma}$}\cdot~\mathbf{Q'}$ from Eqs. (\ref{08})-(\ref{11}). Besides, the r.h.s. of the Eq. (\ref{11}) becomes $Q_t\,\Delta$ and the source equation reduces to 
$$
\ro'''+(\ga_{x}+\ga_{m}+\ga_{r})\ro''+(\ga_{x}\ga_{r}+\ga_{x}\ga_{m}+\ga_{m}\ga_{r})\ro'+
$$
\be
\n{sef}
\ga_{x}\ga_{m}\ga_{r}\ro=Q_t\,\Delta.
\ee

Once the transversal interaction $\mathbf{Q_t}$ is specified we obtain the energy density $\ro$ by solving  the source equation (\ref{sef}) and the component energy densities $\ro_x$, $\ro_m$, and $\ro_r$ after inserting $\ro$ into Eqs. (\ref{08})-(\ref{10}).  We center in a model where the energy exchange is given by transversal interactions which are linearly  dependent on $\ro_{x}$, $\ro_{m}$, $\ro_{r}$, along with their derivatives up to first order, and $\ro$, $\ro'$, $\ro''$, $\ro'''$. 
Hence, after using Eqs. (\ref{08})-(\ref{10}) one finds that 
\be
\label{ql}
Q_{t}=\beta_{1}\ro+\beta_2\ro'+\beta_{3}\ro''+\beta_{4}\ro''',
\ee
becomes a linear functional of the basis elements  $\ro$, $\ro'$, $\ro''$, $\ro'''$ (see \cite{jefe1}, \cite{jefe2}), and $\beta_{1}$, $\beta_2$, $\beta_{3}$, $\beta_{4}$ are four constant parameters. For the interation (\ref{ql}) the source equation (\ref{sef}) turns into a linear differential equation whose general  exact solution reads,
\be
\n{DenZ}
\ro=3H^{2}_{0}\left[{\cal A}(z+1)^{3\ga_{s}}+{\cal B}(z+1)^{3\ga_{+}}+{\cal C}(z+1)^{3\ga_{-}}\right],
\ee
where $a^{-1}=z+1$ and  $z$ is the cosmological redshift while $(\ga_s,\ga_-,\ga_+)$ are the characteristic roots of the source equation (\ref{sef}), and $\ga_s$ is the minimum of $\{\ga_s,\ga_-,\ga_+\}$. The  $\ro_i$ are obtained by inserting Eq. (\ref{DenZ}) into Eqs. (\ref{08})-(\ref{10}) whereas the constants ${\cal A}$, ${\cal B}$ and ${\cal C}$ are written in term of the density parameters  $\Omega_{0i}=\ro_{0i}/3H^{2}_{0}$ which fulfill the condition $\Omega_{x0}+\Omega_{r0}+\Omega_{m0}=1$ for a flat FRW Universe. Additionally, we will choose  $(\ga_{x}, \ga_{m}, \ga_{r})=(0,1,4/3)$ to recover the three self-conserved cosmic components in the limit of vanishing interaction. 

%%%%%%%%%%%%%%%%%%%%%%%%%%%%%%%%%%%%%%%%%%%%%%%%%%%%%%%%%%%%%%%
\section{ Model II:  Matter Coupled with a scalar field and decoupled radiation}
%%%%%%%%%%%%%%%%%%%%%%%%%%%%%%%%%%%%%%%%%%%%%%%%%%%%%%%%%%%%%%% 
 Now, we consider a model with matter $\ro_{\rm m}$ coupled to a scalar field $\ro_\phi$ plus a decoupled component $\ro_{\rm r}$, so that the evolution of the FRW universe is governed by 
\be
\n{01'}
3H^{2}=\ro_{\rm t}=\ro_{\rm m}+\frac{1}{2}\dot\phi^2+V(\phi)+\ro_{\rm r},
\ee
\be
\n{02'}
\dot{\ro}+3H(\ro_{\rm m}+\dot\phi^2)=0,\,\,\,\,\dot\ro_{\rm r}+3H\ga_{\rm r}\ro_{\rm r}=0,
\ee
\bn{ro}
\ro=\ro_{\rm m}+\frac{1}{2}\dot\phi^2+V(\phi),
\ee
where $\ro$ includes all dark components, so $\ro_{\rm t}=\ro+\ro_{\rm r}$. We identify the scalar field with a fluid having  energy density  $\rho_\phi=\frac{1}{2}\dot\phi^2+V(\phi)$, pressure $p_\phi=\frac{1}{2}\dot\phi^2-V(\phi)$ and describe them as a mix of two fluids, namely, $\rho_{\rm sm}=\dot\phi^2/2$ and $\rho_{\rm v}=V(\phi)$, with equations of state $p_{\rm sm}=\rho_{\rm sm}$ and $p_{\rm v}=-\rho_{\rm v}$ respectively, so  $\gamma_{\rm v}=0$, $\ga_{\rm m}=1$, $\gamma_{\rm sm}=2$ and $\ga_{\rm r}$ will be estimated later on. Then, $\rho_{\rm sm}$ can be associated with  stiff matter, $\rho_{\rm v}$ plays the role of vacuum energy, while $\ro_{\rm m}$ represents a pressureless matter component  . 

 When additionally the matter  interacts with the scalar field, we split the Eq. (\ref{02'}) into three balance equations
\be
\n{04'}
\ro_{\rm v}'  =  Q_{\rm v},~~~~~~\ro_{\rm m}' + \ro_{\rm m} = Q_{\rm m},
\ee
\be
\n{06'}
\ro_{\rm sm}' + 2\ro_{\rm sm} = Q_{\rm sm},
\ee
so that the whole conservation equation reads
\be
\n{03'}
\ro'=-\ro_{\rm m}-2\ro_{\rm sm},
\ee
 while the dynamics of the scalar field is given by the following  modified Klein-Gordon (MKG) equation, 
\be
\label{re2}
\ddot{\phi}+3H\dot{\phi}+  V'(\phi)=\frac{-3HQ_{\rm m}}{\dot{\phi}}.
\ee
Our goal is to build a realistic cosmological model with an early radiation-dominated era. Next, the Universe entered  into dark matter era, the radiation becomes subdominant and then follows a dark energy dominated stage at late times. To this end and taking into account that $\mbox{\boldmath ${\gamma}$}=(0,1,2)$,  we adopt the transversal interaction
\bn{qt}
\mathbf{Q_t}=(1,-2,1)Q_{\rm v}.
\ee
By combining Eqs. (\ref{04'}) and (\ref{qt}), we get  $Q_{\rm m}=-2Q_{\rm v}=-2\ro_{\rm v}'=-2(dV/d\phi)\,(d\phi/d\eta)=-2(dV/d\phi)\,(d\phi/3Hdt)$. Thus, the MKG equation (\ref{re2}) becomes $\ddot{\phi}+3H\dot{\phi}- V'(\phi)=0$, showing that $\mathbf{Q_t}$ changes the sign of $ V'(\phi)$ in the KG equation.  We actually get the transversal interaction leads to $Q_{\rm v}=V'(a)$, so it  changes the behavior of the scalar field.

We have built an interacting three fluid model with energy trasnfer (\ref{qt}) defined by Eqs. (\ref{ro}), (\ref{03'}), and  $\ro''=\ro_{\rm m}+ 4\ro_{\rm sm}$ obtained after differentiation Eq. (\ref{03'}) and using Eqs. (\ref{04'})-(\ref{06'}) and (\ref{qt}). By solving this system of equations in the $(\ro_{\rm v}, \ro_{\rm m}, \ro_{\rm sm})$-variables, we find  
\be
\n{08'}
\ro_{\rm v}=\frac{1}{2}\left[ 2\ro+3\ro'+ \ro''\right],
\ee
\be
\n{09'}
\ro_{\rm m}=-\left[ 2\ro'+ \ro''\right], ~~~\ro_{\rm sm}=\frac{1}{2}\left[\ro'+\ro''\right].
\ee

Following  Refs. \cite{jefe1}-\cite{jefe2}, we replace Eqs. (\ref{08'})-(\ref{09'})  into Eq. (\ref{04'}) or (\ref{06'}) and get the source equation for $\ro$: 
\be
\n{11'}
\ro'''+3\ro''+2\ro'=2Q_{\rm v}.
\ee
Using the general solution of the Eq. (\ref{11'}), that has the form of (\ref{DenZ}), and inserting it into Eqs. (\ref{08'})-(\ref{09'}), we have $\ro_{\rm v}$, $\ro_{\rm m}$ and $\ro_{\rm sm}$. There, we have made the substitution $\ro_{\rm v}\to\ro_{\rm x}$, $\ro_{\rm m}\to\ro_{\rm c}$, and  $\ro_{\rm sm}\to\ro_{\rm dr} $  in order to show explicitly the interaction between these components. Then, $\ro_{\rm x}$,  $\ro_{\rm c}$, and $\ro_{\rm dr}$ will represent the dark energy, dark matter and dark radiation energy densities and, we will assume that $\ga_{s}<2/3<\ga_{-}\approx 1<\ga_{+}\approx 4/3$ to show the transition of the universe from an early radiation dominated era to an intermediate dark matter dominated era  (non-baryonic) to end in a dark energy dominated era at late times. Note that the integration constants ${\cal A}$, ${\cal B}$ and ${\cal C}$ can be esxpressed in terms of the density parameters  $\Omega_{\rm v0}=\ro_{\rm v0}/3H^{2}_{0}$, $\Omega_{\rm c 0}=\ro_{\rm c0}/3H^{2}_{0}$, $\Omega_{\rm dr0}=\ro_{\rm dr0}/3H^{2}_{0}$, and  $\Omega_{\rm r 0}=\ro_{\rm r0}/3H^{2}_{0}$ (see \cite{jefe3}). The flatness condition reads $\Omega_{\rm \phi 0}+\Omega_{\rm c0}+ \Omega_{\rm r0} =1$ along with  $\Omega_{\rm v0}+\Omega_{\rm dr0}=\Omega_{\rm \phi 0}$. 

We will now reconstruct the potential as a function of the scalar field $\phi$ by using that $V(a)$, $\dot\phi^2(a)$ and $\ro_t(a)$ are explicit function of the scale factor through $V=\ro_{\rm x}$, $\dot\phi^2/2=\rho_{\rm dr}$ and $\ro_{\rm t}=\dot\phi^2/2+V+\ro_{\rm m}+\ro_{r0}/a^{3\ga_{\rm r}}$. Rewritten $\dot\phi^2/2$ as $\dot\phi^2=3\ro_t\phi'^2$, we easily obtain $\phi=\phi(a)$,
\bn{re3}
\phi=\int{\frac{\sqrt{6\Om_{\rm sm}}}{a}\,\,da},
\ee
where $\Om_{\rm sm}=\dot\phi^2/2\ro_{\rm t}$. After integrate the Eq. (\ref{re3}), we find $\phi(a)$ and then follows $V(\phi)=\ro_{\rm x}(a(\phi))$. Also, the scalar field  energy density $\ro_\phi=V+\dot\phi^2/2=\ro_{\rm x}+\rho_{\rm dr}$ is easily calculated as a real function of the scale factor, 
$$
\ro_{\phi}=3H^{2}_{0}[(\ga_s-1)^2{\cal A}x^{3\ga_{s}} +(\ga_+ -1)^2{\cal B}x^{3\ga_{+}}
$$
\bn{roo}
+(\ga_- -1)^2{\cal C}x^{3\ga_{-}}].
\ee
We have changed the original degree of freedom $\phi$ by the scale factor $a$; such fact  was essential to facilitate the reconstruction process and to examine the observational constraints through the Hubble expansion rate.

In the early radiation-dominated era, where $\ga\simeq \ga_+\approx 4/3$, and in the dark energy dominated era where the universe has an accelerated expansion implying  $\ga\simeq\ga_s$, we obtain the  approximate potential and scalar field, 
\bn{pr}
V\simeq\frac{3H^{2}_{0}}{2}(\ga_+-2)(\ga_+-1)\,{\cal B}\,a^{-3\ga_{+}},
\ee
\bn{fr}
\Delta\phi\simeq-\sqrt{3 \ga_{+}(\ga_{+}-1)}\ln a,
\ee
the former being negative definite and the latter becomes real, while $a\approx t^{2/3\ga_+}$. Hence after having reconstructed the potential and the interaction term $Q_{\rm m}$, we obtain 
\bn{vrr}
V\simeq \frac{3H^{2}_{0}}{2}(\ga_+-2)(\ga_+-1)\,{\cal B}\,e^{\sqrt{3 \ga_{+}/ (\ga_{+}-1)}\,\Delta\phi},
\ee
\bn{qm}
Q_{\rm m}\simeq-3H^{2}_{0}\ga_+(\ga_+-2)(\ga_+-1)\,{\cal B}\,e^{\sqrt{3 \ga_{+}/ (\ga_{+}-1)}\,\Delta\phi},
\ee
in the early epochs.

At late times, when the dark energy  governs the dynamic of the universe,  we find
\bn{ps}
V\simeq \frac{3H^{2}_{0}}{{2}}(\ga_s-2)(\ga_s-1)\,{\cal A}\,a^{-3\ga_{s}},
\ee
\bn{ppf}
\Delta\phi\simeq\sqrt{3\ga_{s}(\ga_{s}-1)}\ln a, 
\ee
hence, after  having applied the reconstruction process, we find that the scalar field is driven by   
\bn{psr}
V \simeq \frac{3H^{2}_{0}}{{2}}(\ga_s-2)(\ga_s-1)\,{\cal A}\, e^{ -\sqrt{3 \ga_{s}/ (\ga_{s}-1 )}\,\Delta\phi},
\ee
a positive potential because $\ga\approx \ga_s$. In this late regime $a\approx t^{2/3\ga_s}$ and the  approximate $Q_{\rm m}\simeq -3H^{2}_{0}\ga_s(\ga_s-2)(\ga_s-1)\,{\cal A}\,e^{\sqrt{3 \ga_{s}/ (\ga_{s}-1)}\,\Delta\phi}.$ 
%%%%%%%%%%%%%%%%%%%%%%%%%%%%%%%%%%%%%%%%%%%%%%%%%%%%%%%%%%%%%%%%%%%%%%%%%%%%%%%%
\section{Observational constraints on a transversal interacting  model: Summary and Discussion}
%%%%%%%%%%%%%%%%%%%%%%%%%%%%%%%%%%%%%%%%%%%%%%%%%%%%%%%%%%%%%%%%%%%%%%%%%%%%%%%%
We will provide a qualitative estimation of the cosmological paramaters by constraining them with the Hubble data  \cite{obs3}- \cite{obs4} and the stricts bounds for the behavior of dark energy at early times \cite{EDE1}-\cite{EDE2}. In the former case, the  statistical analysis is based on the $\chi^{2}$--function of the Hubble data \cite{obs3}, \cite{obs4}. For model I,  the Hubble expansion is given by
\be
\n{Ht}
H_{\rm I}(\theta| z)=H_{0} \Big( {\cal A}x^{3\ga_{s}}+{\cal B}x^{4}+{\cal C}x^{3}\Big)^{\frac{1}{2}}.
\ee
Fron now on  we will  choose  $\ga_{+}=4/3$ and  $\ga_{-}=1$, so the parameters of the model are   $\theta=\{H_{0},\ga_{s},  \Omega_{x0},\Omega_{m0}\}$, and  will use that the flatness condition implies $\Omega_{r0}=1-\Omega_{x0}-\Omega_{m0}$.
\begin{figure}[hbt!]
\begin{minipage}{1\linewidth}
\resizebox{1.6in}{!}{\includegraphics{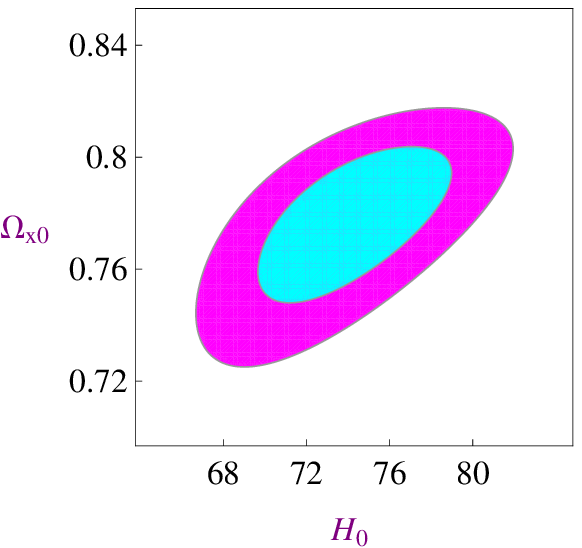}}
\resizebox{1.6in}{!}{\includegraphics{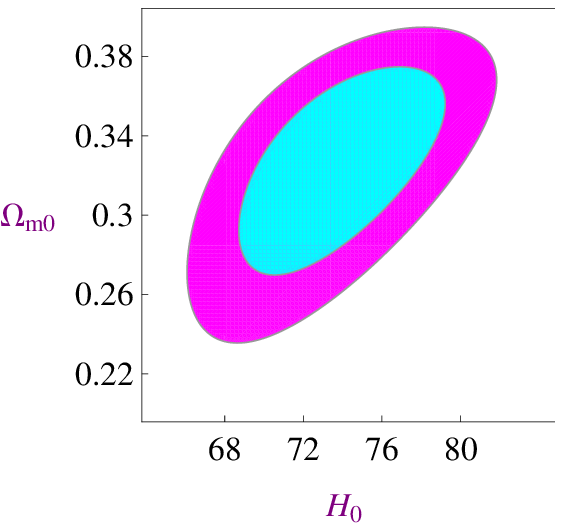}}\hskip0.05cm
\resizebox{1.6in}{!}{\includegraphics{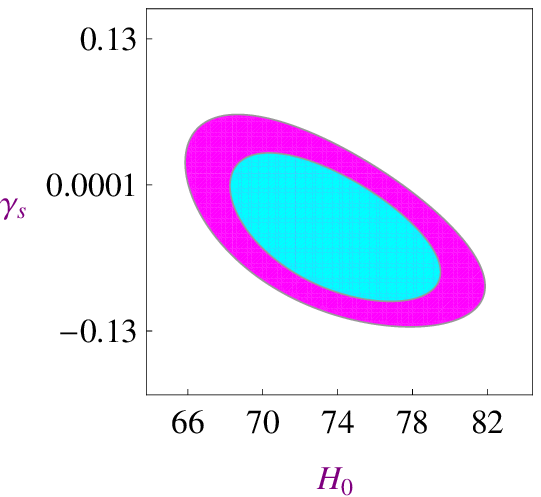}}\hskip0.05cm 
\resizebox{1.6in}{!}{\includegraphics{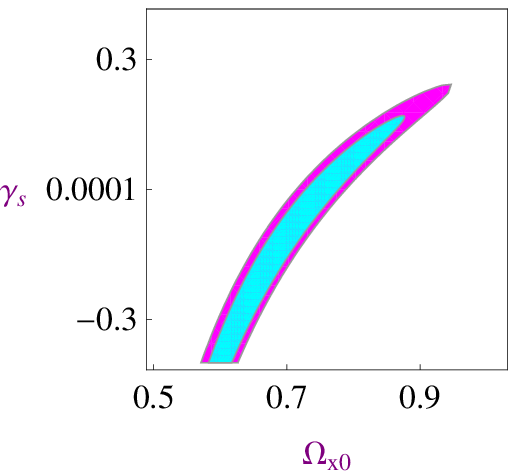}}\hskip0.05cm 
%\resizebox{1.6in}{!}{\includegraphics{6.eps}}
%\resizebox{1.6in}{!}{\includegraphics{7.eps}}
\caption{\scriptsize{Model I: two-dimensional C.L. associated with $1\sigma$,$2\sigma$ for different $\theta$ planes.}}
\label{Fig1}
\end{minipage}
\end{figure}
We obtain that
$\ga_{s}$ varies from $10^{-4}$ to $10^{-3}$, so  these values clearly fulfill
the constraint $\ga_{s}<2/3$ that assure the existence of accelerated phase of
the universe at late times [see Fig. (\ref{Fig1})]. We find the best-fit at 
$(H_{0}, \Omega_{x0})=(74.32 {\rm km~s^{-1}\,Mpc^{-1}},0.77)$ with $\chi^{2}_{\rm d.o.f}=0.783$
by using the priors $\Omega_{m0}=0.2$ and $\ga_{s}=10^{-3}$.  The value of $\Omega_{x0}$
is slightly greater than the standard one of $0.7$ being such discrepancy only of $0.1\%$.
Moreover, we find that using the priors  $(H_{0}, \ga_{s})=(74.20 {\rm km~s^{-1}\,Mpc^{-1}}, 10^{-3})$
the best-fit values for the present-day density parameters  are 
$(\Omega_{x0},\Omega_{m0})=(0.74, 0.23)$ along with a lower goodness condition ($\chi^{2}_{\rm d.o.f}=0.779$). 
In performing the statistical analysis, we find that $H_{0} \in [71.28, 74.32]{\rm km~s^{-1}\,Mpc^{-1}}$ 
so the estimated values are met  within $1 \sigma$ C.L.  reported
by Riess \emph{et al}, to wit,  $H_{0}=(72.2 \pm 3.6){\rm km~s^{-1}\,Mpc^{-1}}$.  We also find that the $z_{t}$ (transition redshift) is of the order unity varying over the interval $[0.63, 0.75]$, such values are close 
to $z_{t}=0.69^{+0.20}_{-0.13}$ reported in \cite{Zt1}-\cite{Ztn}.
 
\begin{figure}[hbt!]
\begin{minipage}{1\linewidth}
\resizebox{1.5in}{!}{\includegraphics{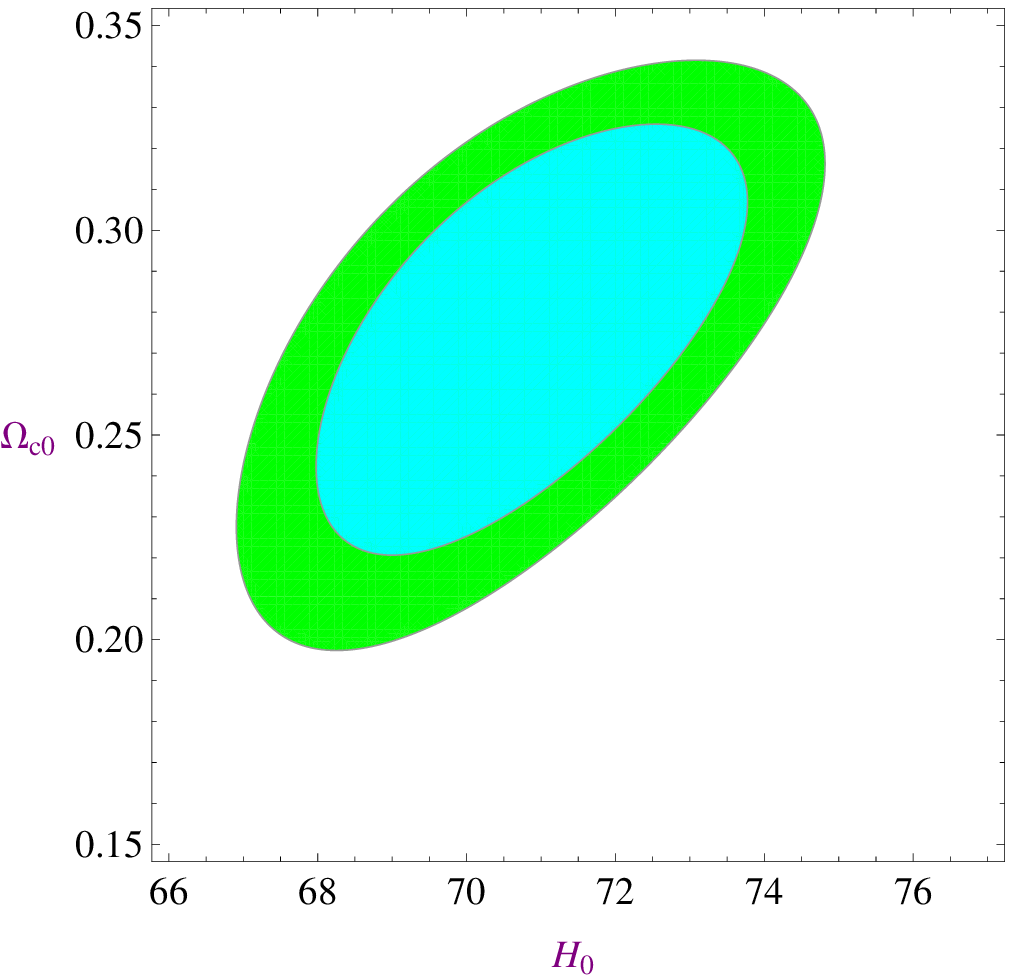}}\hskip0.05cm
\resizebox{1.5in}{!}{\includegraphics{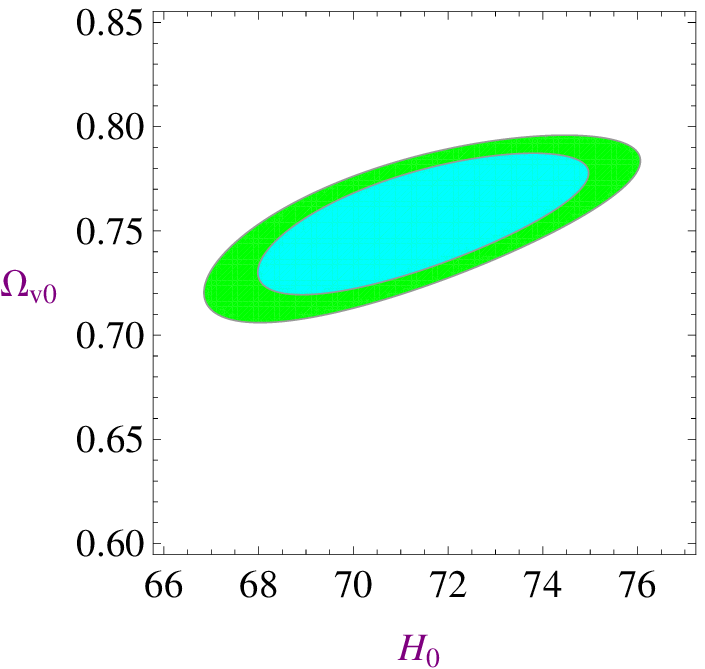}}\hskip0.05cm
\resizebox{1.5in}{!}{\includegraphics{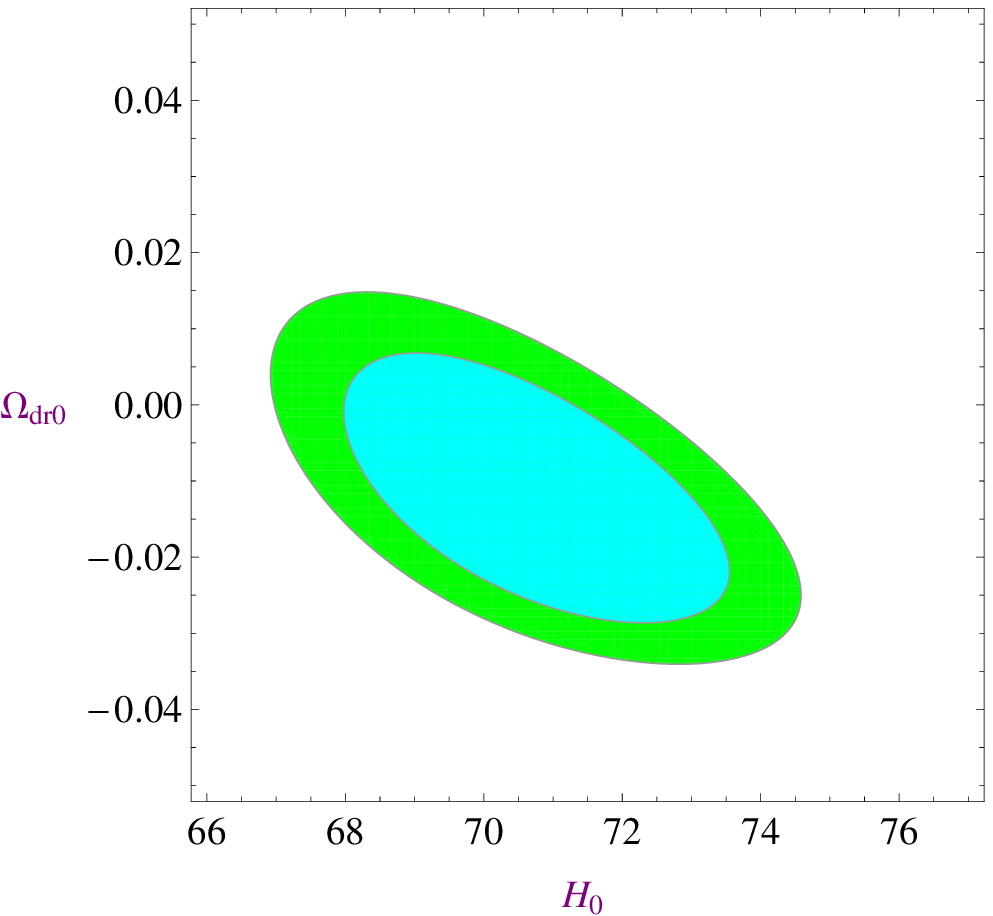}}\hskip0.05cm 
\resizebox{1.5in}{!}{\includegraphics{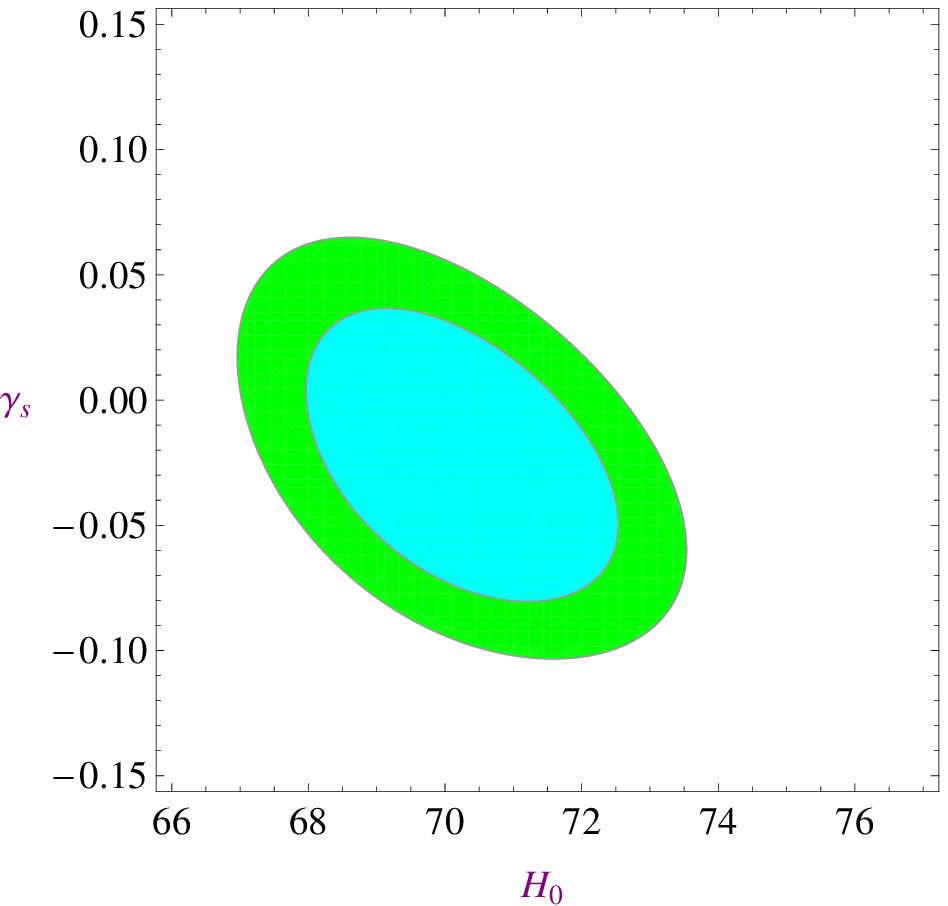}}\hskip0.05cm 
%\resizebox{1.5in}{!}{\includegraphics{c5.eps}}\hskip0.05cm
%\resizebox{1.5in}{!}{\includegraphics{c6.eps}}\hskip0.05cm
%\resizebox{1.5in}{!}{\includegraphics{c7.eps}}\hskip0.05cm
%\resizebox{1.5in}{!}{\includegraphics{c8.eps}}\hskip0.05cm
%\resizebox{1.5in}{!}{\includegraphics{c9.eps}}\hskip0.05cm
%\resizebox{1.5in}{!}{\includegraphics{c10.eps}}\hskip0.05cm
%\resizebox{1.5in}{!}{\includegraphics{c11.eps}}\hskip0.05cm
%\resizebox{1.5in}{!}{\includegraphics{c12.eps}}\hskip0.05cm
\caption{\scriptsize{Model II: two-dimensional C.L. associated with $1\sigma$,$2\sigma$ for different $\theta$ planes.}}
\label{Fig2}
\end{minipage}
\end{figure}

In the case of  model II, one finds that the Hubble expansion takes the following form:
\be
\n{Ht}
H_{\rm II}(\theta| z)=H_{0} \Big( {\cal A}x^{3\ga_{s}}+{\cal B}x^{3\ga_{+}}+{\cal C}x^{3\ga_{-}}+ {\cal D}x^{3\ga_{r}}\Big)^{\frac{1}{2}}.
\ee
Here, we consider  $\theta=\{H_{0},\ga_{\rm s},  \ga_{+}, \ga_{-},\ga_{\rm r}, \Omega_{\rm  v0},\Omega_{\rm dr0}, \Omega_{\rm c0}\}$ as the independent parameters to be constrained  for the model encoded in the Hubble function (\ref{Ht}).  We start our statistical estimations by performing a global analysis on the eight parameters that characterize the model. In doing so, we find  that $\chi^{2}(\theta)$ reaches a minimum at  $\{H_{0},\ga_{\rm s},\ga_{+},\ga_{-},\ga_{\rm r},\Omega_{\rm  v0},\Omega_{\rm dr0},\Omega_{\rm c0}\}=(70.00, 0.010, 1.300, 1.009, 1.3400, 0.749, 0.00006, 0.199  )$ along with  $\chi^{2}_{\rm d.o.f}=20.172/(29-8)\simeq 0.9605 <1$. We see that
$\ga_{s}$ varies from $ 0.010^{+0.007+0.055}_{-0.090-0.112}$ to $ 0.010^{+0.271+0.335}_{-0.189-0.303}$ at $1\sigma-2\sigma$ confidence levels, so  these values fulfill
the constraint $\ga_{s}<2/3$ at $95\%$ C.L. that ensure the existence of accelerated phase of
the universe at late times [ See Fig. (\ref{Fig2})]. We find the best fit at $ (H_{0},\Omega_{\rm v0})= (70.00^{+4.99+6.07}_{-2.03-3.16} {\rm km~s^{-1}\,Mpc^{-1}}, 0.7299^{+0.057+0.0658}_{-0.0109-0.0238})$ with $\chi^{2}_{\rm d.o.f}=0.736$
by using the priors $(\Omega_{\rm c0}, \Omega_{\rm dr0},\ga_{s},\ga_{+},\ga_{-},\ga_{\rm r})=(0.1999, 6\times 10^{-5}, 0.010, 1.3000, 1.0099, 1.3400)$.  Moreover, we find that using the priors $( H_{0},\Omega_{\rm dr0},\ga_{s},\ga_{+},\ga_{-},\ga_{\rm r})=(70.00 {\rm km~s^{-1}\,Mpc^{-1}}, 6\times 10^{-5}, 0.010, 1.3000, 1.0099, 1.3400)$
the best-fit values for the present-day density parameters are considerably improved, namely, these turn give  
($\Omega_{\rm v0}, \Omega_{\rm c 0})= (0.729^{+0.167+0.218}_{-0.102-0.152}, 0.239^{+0.210+0.355}_{-0.281-0.364})$ along with the same goodness condition ($\chi^{2}_{\rm d.o.f}=0.7368$) .  Notice that $\Omega_{\rm c0}$ varies from
$0.239^{+0.086+0.101}_{-0.019-0.042}$ to $0.239^{+0.215+0.267}_{-0.091-0.115}$ at $68\%$, $95\%$ C.L., whereas $H_{0} \in [70.00^{+2.57+3.54}_{-2.04-3.05}, 70.00^{+4.99+6.07}_{-2.03-3.16}]{\rm km~s^{-1}\,Mpc^{-1}}$; thus,  the latter values are consistent with the analysis of ACT and WMAP-7 data that 
gives $H_0 = 69.7 \pm 2.5{\rm km~s^{-1}\,Mpc^{-1}}$. 
We find that the transition redshift is   $z_{\rm t}=0.604^{+0.203+0.325}_{-0.025-0.043}$ at $1\sigma, 2\sigma$ C.L.,  close to ones reported in \cite{Zt1}, \cite{Ztn} or with the marginalized best-fit values $z_{\rm t}=0.623^{+0.039
}_{- 0.052}$ listed in \cite{mizt}, \cite{mizt2}, \cite{Zt2}.
%The present-day value of decelerating parameter  
% varies as  $q_{0}=-0.579^{+0.350+0.470}_{-0.247-0.354}$ at $68\%$, $95\%$ C.L., so  this estimation  is similar to $q_{0} = -0.53^{+0.17}_{-0.13}$ \cite{Waga} or with the marginalized best-fit values $q_{0}=-0.671^{+0.120}_{-0.283}$ listed in \cite{mizt}, \cite{mizt2}. 

The behavior of dark energy at early times can be considered as a new cosmological probe for testing dynamical dark energy models. The fraction of dark 
energy at recombination epoch should fulfill the bound $\Omega_{\rm ede}(z\simeq 1100)<0.1$ in order to the dark energy model be consistent 
with the big-bang nucleosynthesis (BBN) data.   Some signals could arise from the early dark energy (EDE) models uncovering the nature of DE as well as 
their properties to high redshift, giving an invaluable clue to the physics behind the recent speed up of the universe \cite{EDE1}. Then,  it was examined the current and future data  for constraining the amount  of EDE, the cosmological data analyzed has led to an upper bound of  $\Omega_{\rm ede}(z\simeq 1100)<0.043$ with $95\%$ confidence level (CL) in case of relativistic EDE while for a quintessence type of EDE has given $\Omega_{\rm ede}(z\simeq 1100)<0.024$ although   the EDE component is not preferred, it is also not 
excluded from the current data \cite{EDE1}. Another forecast for  the bounds of the EDE  are obtained with the  Planck  and CMBPol experiments\cite{EDE2}, thus  assuming a $\Omega_{\rm ede}(a \simeq 10^{-3}) \simeq 0.03$  for studying 
the stability of this value, it found that $1\sigma$ error coming from Planck experiment 
is $\sigma^{\rm Planck}_{\rm ede} \simeq 0.004$ whereas the CMBPol  improved this bound by a factor 4 \cite{EDE2}.  

For the model I, we find that at  early times the dark energy  changes rapidly with the redshift $z$ over the interval 
$[10^{3}, 10^{4}]$ , indeed,  around $z \simeq 1100$ we find that 
$2.4 \times 10^{-11} \leq \Omega_{\rm x}  \leq 2.6 \times 10^{-10}$. However for the model II, we find that the dark energy   around $z \simeq 1100$, $\Omega_{\rm \phi} \simeq 0.014$.  In  comparison with the SPT constraints on the early dark energy density over
WMAP7 alone it was found that the $95\%$ upper limit on $\Omega_{\rm ede}$ is reduced from $0.052$ for WMAP7-only to $0.013$ for
WMAP7+SPT. This is a $38\%$ improvement on the upper limit of $\Omega_{\rm ede} < 0.018$ reported for WMAP7+K11 \cite{Reic}. The upper limit is essentially unchanged at $\Omega_{\rm ede} < 0.014$ for WMAP7+SPT+BAO+SNe. The $\Omega_{\rm ede} < 0.013$ bound from WMAP+SPT is the best
published constraint from the CMB (see \cite{Hou} and reference therein). Our findings point out that the model constructed here not only fulfill the severe bound
of $\Omega_{\rm \phi}(z\simeq 1100)<0.018$ obtained from the measurements of CMB anisotropy from ACT and SPT \cite{Reic}, \cite{Amendola}, \cite{Hou}  but also is consistent with the future constraints
achievable by Planck and CMBPol experiments \cite{EDE2} as well, corroborating that the value of the cosmological parameters selected before, 
through the statistical analysis made with Hubble data, are consistent with BBN constraints. Besides,  regarding the values reached by dark energy  around $z=10^{10}$ (BBN), we find that 
 $\Omega_{\rm \phi}=0.0025$ at $1\sigma$level indicating the conventional BBN processes that occurred at temperature of $1 {\rm  Mev}$ is not spoiled because the severe
bound reported  for early dark energy $\Omega_{\rm \phi}(z\simeq 10^{10})<0.21$\cite{Cyburt} or a strong upper limit $\Omega_{\rm \phi}(z\simeq 10^{10})<0.04$ \cite{Cyburt2} are fulfilled at BBN.

\begin{figure}[hbt!]
\begin{minipage}{1\linewidth}
\resizebox{1.5in}{!}{\includegraphics{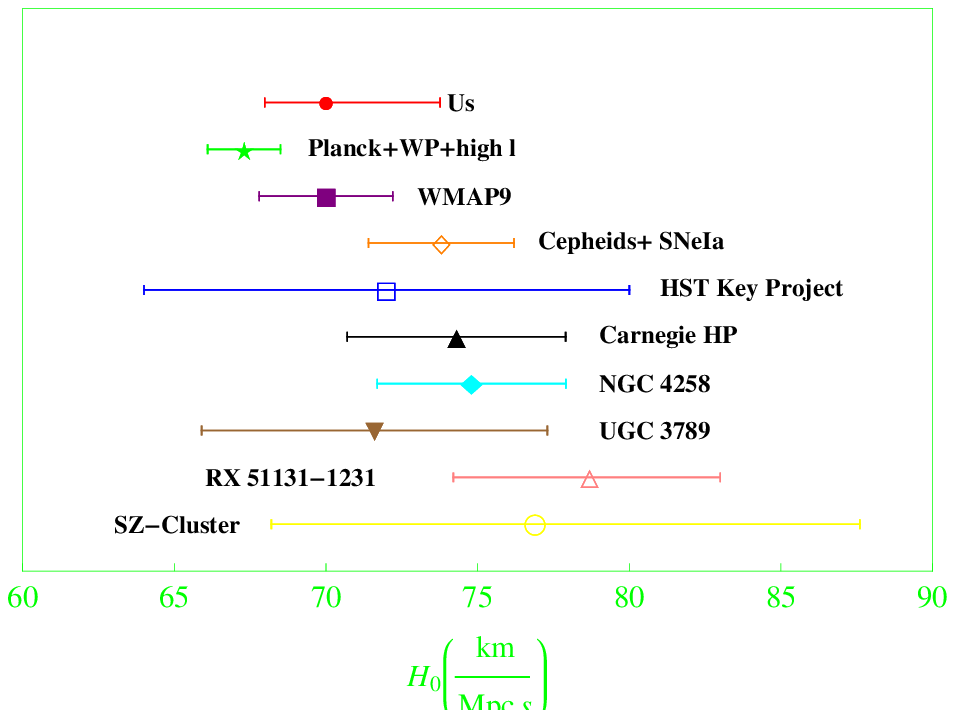}}
\resizebox{1.5in}{!}{\includegraphics{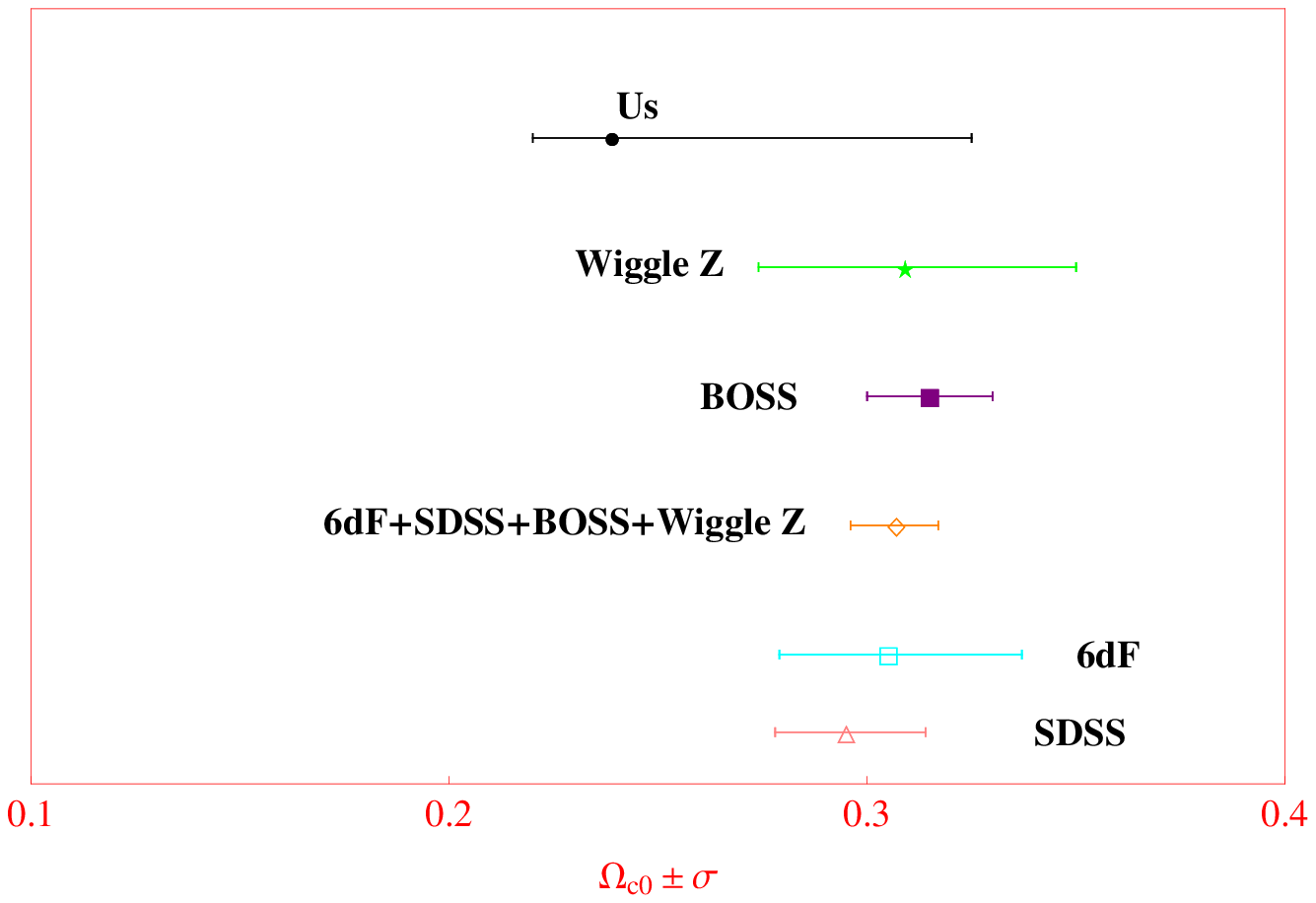}}\hskip0.05cm
\resizebox{1.5in}{!}{\includegraphics{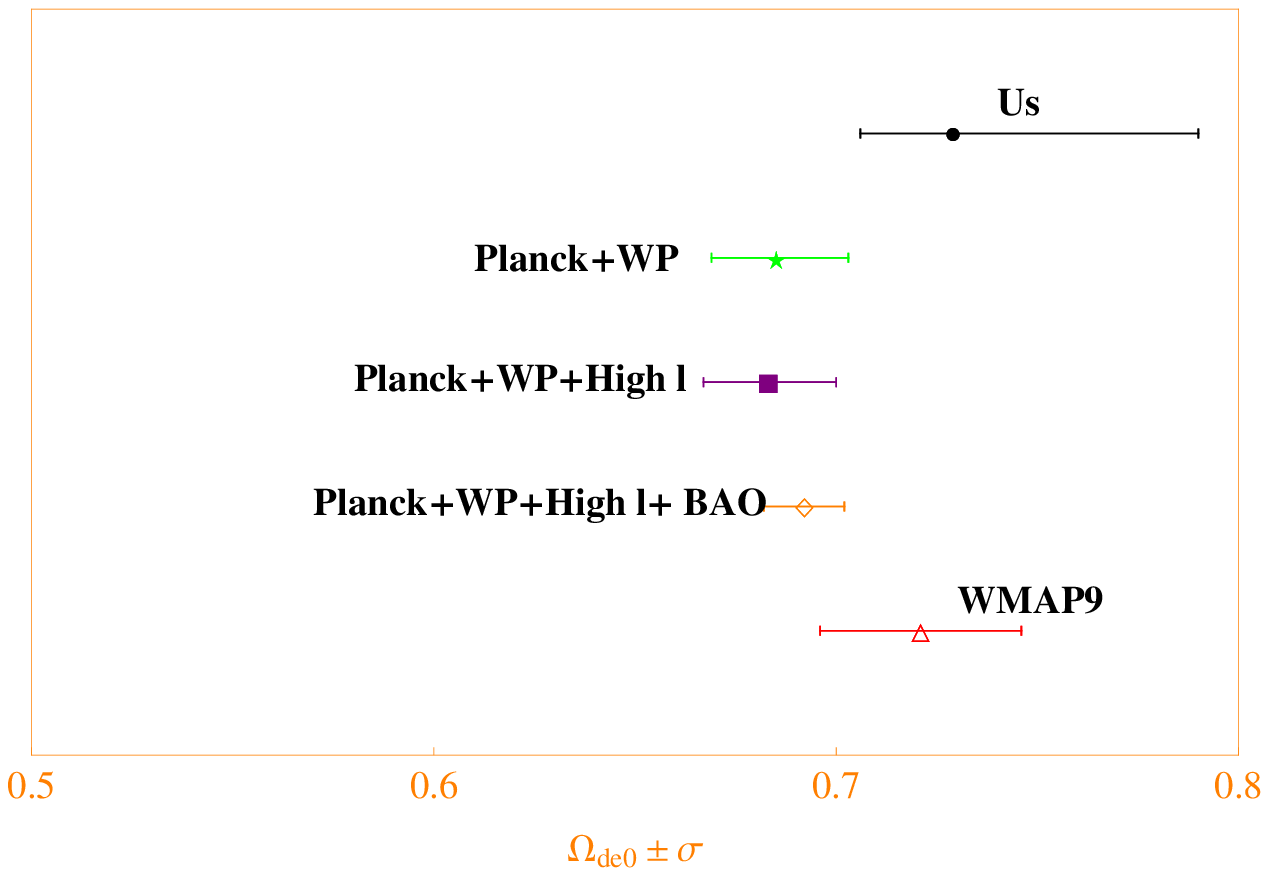}}\hskip0.05cm
\caption{\scriptsize{Comparison of $H_{0}$, $\Omega_{\rm c0}$, and $\Omega_{\rm v0}$ measurements, with estimates of $\pm 1\sigma$ errors, from a number of different  methods. These are compared with the data provided by Planck and WMAP9.}}
\label{H1}
\end{minipage}
\end{figure}

We will perform a further  comparative analysis by taking into account several data sets along with  theirs  constraints on the most relavant parameters \cite{Planck2013}, \cite{WMAP9}, namely, the values  or bounds reported for  $H_{0}$, $\Omega_{\rm v0}$, $\Omega_{\rm c0}$, and $\Omega_{\rm ede}$.  The  Planck power spectra leads to a low value of the Hubble constant, which is
tightly constrained by CMB data alone within the $\Lambda$CDM model. From the Planck+WP+highL analysis, it found that $H_{0}   = (67.3\pm 1.2) {\rm km s^{-1} Mpc^{-1}}$ at $68\%$ \cite{Planck2013}. A low value of $H_{0}$   has been found in other CMB experiments, most notably from the recent WMAP-9 analysis. Fitting
the base $\Lambda$CDM model for the WMAP-9 data, it is found $ H_{0} =(70.0 \pm  2.2) {\rm km s^{-1} Mpc^{-1}} $ at $68\%$ C.L. \cite{WMAP9}. Then, our best estimation $ H_{0} =70.00^{+3.76}_{-2.02} {\rm km s^{-1} Mpc^{-1}} $ at $68\%$ C.L  is perfectly in agreement with the value reported by WMAP-9 project but  shows a slightly difference with the  Planck+WP+highL data, which is less than $0.04\%$. In Fig. (\ref{H1}), we show other observational estimations of the Hubble constant that includes the megamaser-based distance to NGC4258, Cepheid observations, a revised distance to the Large
Magellanic Cloud, and so on (see \cite{Planck2013}). For dark matter,  Wiggle-Z data give $\Omega_{\rm c0}=0.309^{+0.041}_{-0.035}$,  while Boss experiment increases  dark matter amount in $0.019\%$; the joint statistical analysis with  6dF+SDSS+ BOSS+ Wiggle-Z data lead to $\Omega_{\rm c0}=0.307^{+0.010}_{-0.011}$ at $1\sigma$ level \cite{Planck2013}, showing a discrepancy with our  estimation of dark matter, $\Omega_{\rm c0}=0.239^{+0.086}_{-0.019}$,  no bigger than  $0.24\%$  [see Fig. (\ref{H1})].  For the fraction of dark energy,  Planck+WP data give  $0.685 ^{+0.018}_{-0.016}$ at $68\%$ C.L,  whereas the joint statistical analysis on  Planck+WP+highL+ BAO  gives   $0.692\pm 0.010$ at $1\sigma$ level \cite{Planck2013}. In our case,  we find that  $\Omega_{\rm v0}=0.7299^{+0.057}_{-0.0109}$, then the relative difference between Hubble data and Planck+WP+highL is $0.068\%$ [see Fig. (\ref{H1})].  Besides, the CMB anisotropies measurements put further  constraints on the behavior of dynamical dark energy in the recombination epoch; the latest constraints on early dark energy come from  Planck+WP+highL data and leads to $\Omega_{\rm ede} < 0.009$  at $95\%$ C.L \cite{Planck2013}. We have found that $\Omega_{\rm \phi}(z\simeq 10^{3})\simeq 0.014$, so the relative error between both estimations is  $|\Omega^{\rm Us}_{\rm \phi}-\Omega^{\rm Planck}_{\rm ede}|/|\Omega^{\rm Planck}_{\rm ede}| \simeq  0.5 \%$. The latter disagreement can be reduced in a future by including additional measurements of Hubble data points or   other data sets \cite{Xu}, allowing to improve the statistical analysis performed here.

\acknowledgments

L.P.C thanks UBA under Project No. 20020100100147 and CONICET under Project PIP 114-200801-00328 for their partial support. M.G.R is supported by Postdoctoral Fellowship program of  CONICET. 

%%%%%%%%%%%%%%%%%%%%%%%%%%%%%%%%%%%%%%%%%%%%%%%%%%%%%%
%\section{ summary and conclusions}
%%%%%%%%%%%%%%%%%%%%%%%%%%%%%%%%%%%%%%%%%%%%%%%%%%%%%%%%%%%%%%%%%%%%%%

%%%%%%%%%%%%%%%%%%%%%%%%%%%%%%%%%%%%%%%

\end{document}